\documentclass[]{aastex}

\shorttitle{Average flaring chromospheric-coronal density structure from HXR observations}
\shortauthors{Saint-Hilaire et al.}

\begin{document}

\title{Statistically-derived flaring chromospheric-coronal density structure from non-thermal X-ray observations of the Sun}

\author{P. Saint-Hilaire\altaffilmark{1}, S. Krucker\altaffilmark{1,2}, and R.P. Lin\altaffilmark{1,3,4}}
\affil{Space Sciences Laboratory, University of California, Berkeley, CA 94720, USA}
\affil{Institute for 4D Technologies, University of Applied Sciences Northwestern Switzerland, 5210 Windisch, Switzerland}
\affil{Department of Physics, University of California, Berkeley, CA 94720, USA}
\affil{School of Space Research, Kyung Hee University, Yongin, Gyeonggi 446-701, Korea}

\email{shilaire@ssl.berkeley.edu}

\begin{abstract}

	For the first time, we have used RHESSI's spatial and energy resolution to determine the combined chromospheric and coronal density profile of the flaring solar atmosphere in a statistical manner,
	using a dataset of 838 flares observable in hard X-rays above 25 keV.
	Assuming the thick-target beam model, our ``average flaring atmosphere'' was found to have density scale heights of 131$\pm$16 km at low altitudes (chromosphere, up to $\approx$1--1.5 Mm above photosphere), and of 5--6 Mm at high altitudes (corona, above $\approx$2--3 Mm).
	Assuming a unit step change in ionization level, modeling yields a height of 1.3$\pm$0.2 Mm for the transition between fully neutral to fully-ionized atmosphere.
	Furthermore, centroids of emission above 50 keV, produced by electrons of similar or higher energies, are located mostly in a small region $\sim$0.5 Mm in vertical extent, where neutral densities are beyond 3$\times$10$^{13}$ cm$^{-3}$.

\end{abstract}

\keywords{Sun: flares -- Sun: particle emission -- Sun: X-rays -- Sun: Corona}

\section{Introduction}
	
	The density structure of the Sun's atmosphere in the vicinity of the transition region is not well known, at least for flares.
	Most of our knowledge is derived from empirical models \citep[see e.g.][]{VAL1981,Fontenla1993,Gabriel1976,Ewell1993}.

	More recently, \citet{Aschwanden2002,Liu2006,Kontar2008a,Prato2009}, have attempted to derive the chromospheric density structure with the use of hard X-ray (HXR) emission from flare footpoints at different energies, assuming the thick-target beam model \citep{Brown1971}.
	To fit their data, \citet{Aschwanden2002} have assumed that the density has a power-law shape with altitude above photosphere, whereas \citet{Kontar2008a}, working at slightly higher energies (and hence, deeper in the chromosphere), have assumed an exponential shape.
	Uniform target ionization \citep{Brown1973,Kontar2002} was assumed in both case (fully-ionized for the former, fully neutral for the latter).
	The Caltech Irreference Chromospheric Model \citep[CICM,][]{Ewell1993} supports a two-exponential atmosphere, with the lower component's scale height closely corresponding to the one derived by \citet{Kontar2008a},
	and to the \citet[][thereafter VAL]{VAL1981} and \citet[][thereafter FAL]{Fontenla1993} models (i.e. $\sim$130 km).

	These previous studies used single events.
	\citet{Matsushita1992} and \citet{Sato2006} have statistically derived the altitude difference between spatially-averaged HXR emissions at different energies using {\it Yohkoh} HXT's \citep{Kosugi1992} four channels, covering an energy range between 14 to 93 keV.
	We used the same technique of determining spatially-averaged centroids (center of mass) at different energies, and in a similar energy range, but with flares observed by the Ramaty High Energy Solar Spectroscopic Imager \citep[RHESSI,][]{Lin2002}, which has a much higher spectral resolution ($\sim$1 keV).
	This will allow us to carry the data analysis one step further, deducing densities from the regions where non-thermal HXR emission is observed, assuming a thick-target beam model for bremsstrahlung emission.
	
	We will first derive densities using a simple, direct method, and then try to fit a double-exponential with unit step ionization change density model.
\section{Observations} \label{sect:obs}
		
\subsection{Data selection}

	We have selected all flares in the RHESSI flare list\footnote{http://hesperia.nasa.gov/hessidata/dbase/hessi\_flare\_list.txt}, between the start of the RHESSI mission on 2002 February 5 and 2010 January 1.
	Accompanying the flare list are ``Quicklook Images'' in different energy bands (3--6, 6--12, 12--25, 25--50, 50--100, and 100--300 keV), typically accumulated over two minutes.
	Quicklook Images in a certain energy band are made only if they were deemed to ``reliably image'' the X-ray source, that is, 
	that an X-ray source was observed to be roughly at the same location in a majority of RHESSI sub-collimators.
	We examined all events for which such images above 25 keV existed. There were 838 of them.
	
\subsection{Derivation of altitude-energy relationship}

	We assume that non-thermal electrons precipitate along magnetic field lines that radially extend from the photosphere to the corona.
	They propagate from an acceleration region somewhere in the corona, and lose energy and emit bremsstrahlung HXR as per the thick-target model \citep[e.g.][]{Brown2002}.
	Under these assumptions, the difference in source altitude, ${\Delta}h_i$, between energy $\varepsilon_i$ and a reference energy $\varepsilon_{ref}$ can be used to approximate the average density between these two points \citep[see e.g.][and Appendix \ref{appendix:directderivation}]{Brown2002}:
	The idea is that for an accelerated electron distribution with a negative power-law index propagating towards region of higher densities, 
	non-thermal emission at energy $\varepsilon$ spatially peaks where electrons have crossed a column density $N \approx \frac{\varepsilon^2}{2K}$, with $K$ a constant.
	Hence, knowing the distance $s$ between peaks at emission $\varepsilon_1$ and $\varepsilon_2$, one can get an average density between the two peaks of emission: $n=\frac{N_2-N_1}{s}=\frac{\varepsilon_2^2-\varepsilon_1^2}{2Ks}$.
	In actuality, $N = \frac{\varepsilon^2}{2K}$ is only a handy approximation: as discussed in \citet{Brown2002} \citep[see also][]{Xu2008}, 
	there is a dependence on the spectral distribution of injected electrons, and also on the density profile of the medium through which the electron beam is propagating. 
	While both dependences can easily be taken into account when using forward-fitting techniques on singular events, both vary from flare to flare, 
	and we have thus decided to use the $N = \frac{\varepsilon^2}{2K}$ relationship throughout our statistical study.

	We have computed the RHESSI visibilities \citep{Hurford2002}, accumulated over three minutes around peak HXR flux, in the following energy bands: 6--10, 10--15, 15--20, 20--25, 25--30, 30--35, 35--40, 40--50, 50--60, 60--70, 70--80, 80--90, 90--100.
	We have used the hsi\_vis\_fwdfit.pro routine to find positions of centroid at different energies, using subcollimators (SCs) 3--9.
	This method basically yields the position of the centroid of the flux (i.e. its center of gravity).

	The software also yields error bars on centroid positions.
	Typically, higher energies, having less count statistics, yield larger error bars.
	We omitted cases where the software did not converge to a solution.
		
	Once the centroid positions are determined at all energy bands and for all flares, we statistically determine the altitude difference, $\Delta{h}_i$,
	between emission at all energies $\varepsilon_i$ and a reference energy $\varepsilon_{ref}$.
	The $\Delta{h}_i$ are derived using the same method as in \citet{Sato2006} (and described in more mathematical details in Appendix~\ref{appendix:dh}).	
	Figure~\ref{fig:rvsdr} shows plots of $R$ vs. $\Delta{R}$, with $R$ the projected (on the usual plane of the Sun, perpendicular to the observer's line of sight) distance between HXR emission and Sun center, 
	and $\Delta{R}$ the projected distance between the centroid of emissions at different energies.	
	The slope $\frac{\Delta{R}}{R}$ is obtained by linear fitting {\it with a zero intercept coordinate} (i.e. going through the origin).
	Because high energies have lesser statistics, leading to poorer positional accuracy of the centroid, 
	and because there is no potential contamination in the determination of non-thermal centroid position by thermal emission below 35 keV, 
	(although other effects exist and are discussed in Section~\ref{sect:contamination}), the reference energy was chosen to be 35 keV throughout.
	Note that there are many outliers, but that their error bars are generally larger, thus have a lesser contribution to the determination of $\frac{\Delta R}{R}$.
	Finally $\Delta{h}_i$=$r_s \frac{\Delta{R}}{R}$, with $r_s$=solar radius (see Appendix~\ref{appendix:dh}).

		\begin{figure*}[ht!]
		\centering
		\includegraphics[width=12cm]{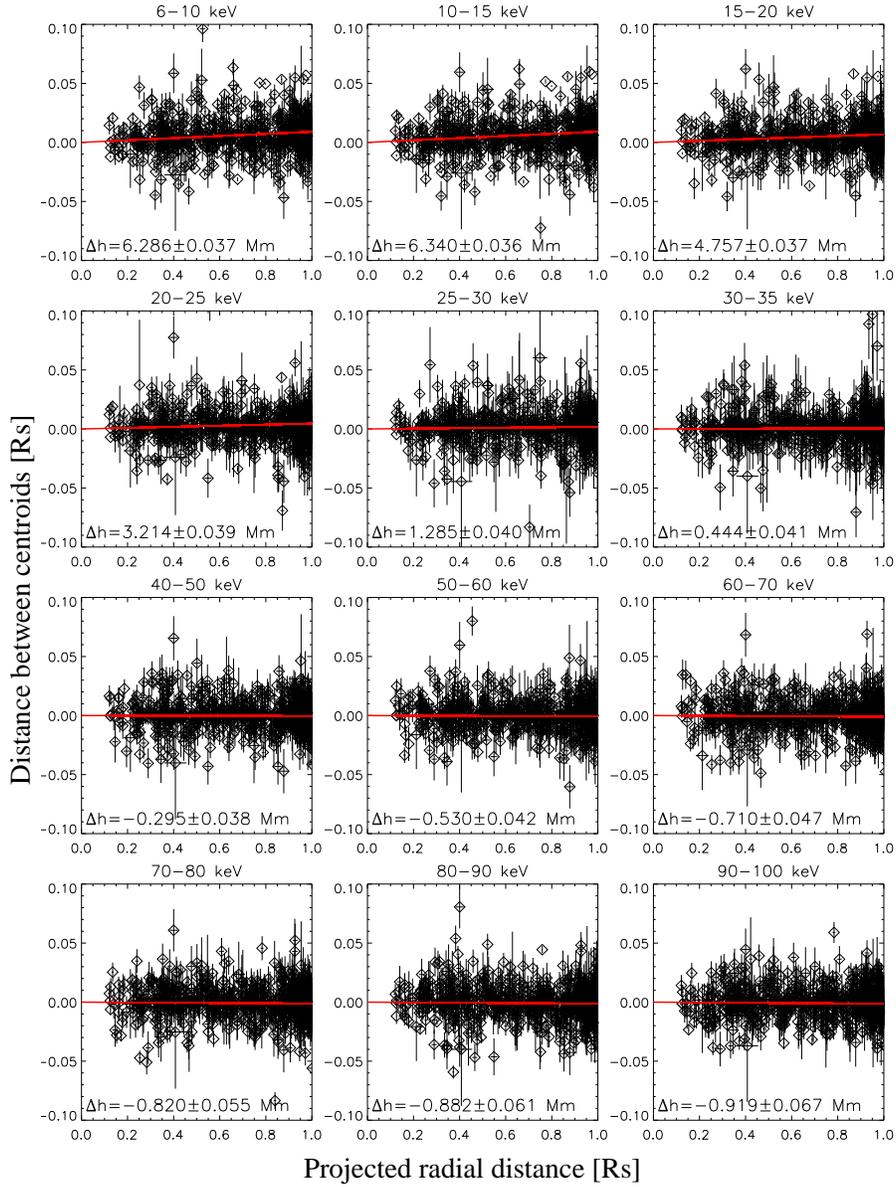}
		\caption{
			{\it Abscissa:} $R$ the projected distance between Sun center and flare emission, 
			{\it ordinates:} ${\Delta}R$, the projected distance between emission centroid at 35--40 keV and at the energy band indicated on top of each plot, with error bars.
		}
		\label{fig:rvsdr}
		\end{figure*}

\begin{figure}[ht!]
		\centering
		\includegraphics[width=7.5cm]{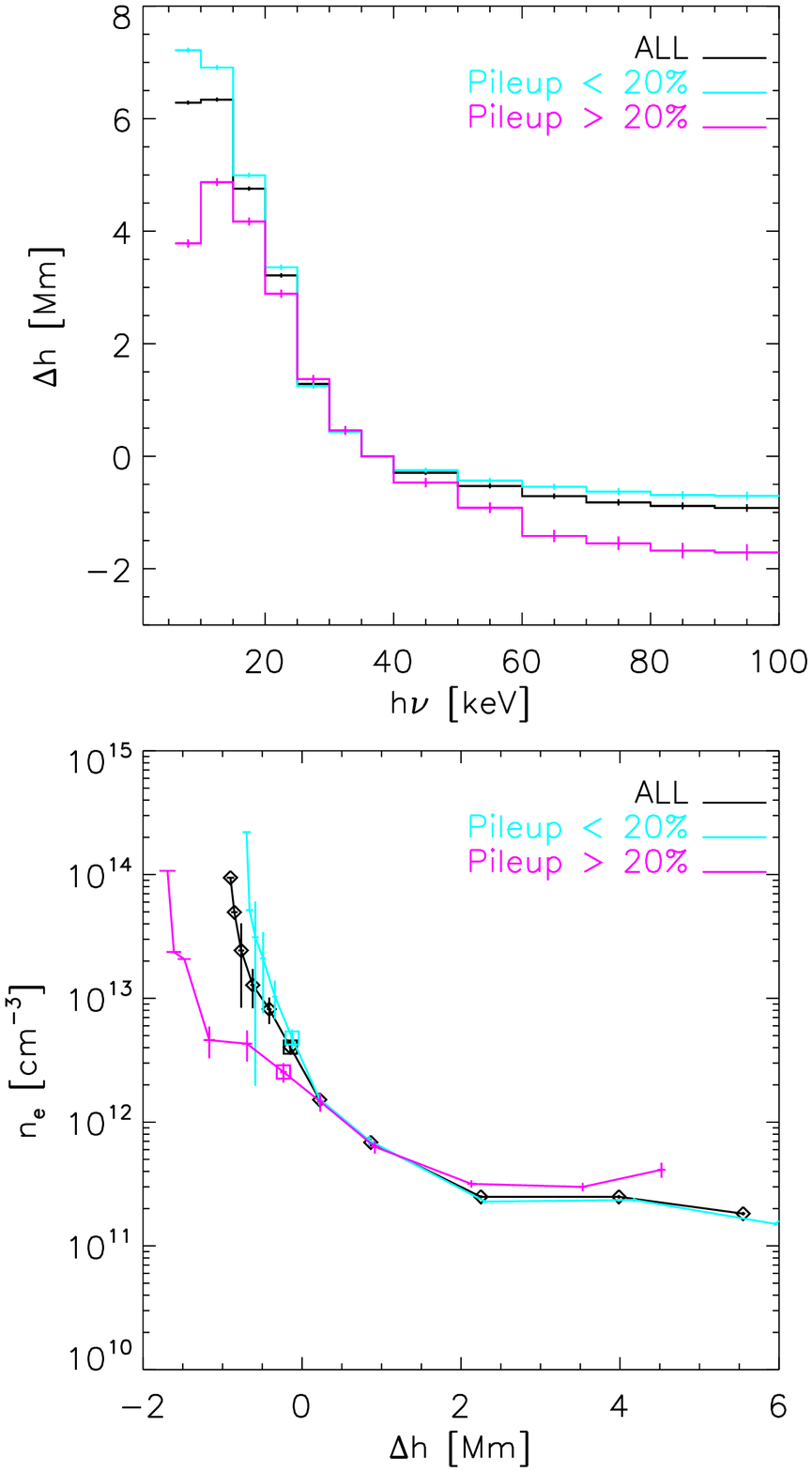}
		\caption{
			{\bf Top:} Plot of $\Delta{h}$=$h-h_{ref}$ as a function of energy $\varepsilon=h\nu$, with $h_{ref}$ being the unknown altitude of peak reference energy $\varepsilon_{ref}$ emission.
			$\varepsilon_{ref}$ was chosen to be 35 keV (see text for more details).
			The error bars were propagated from the $\frac{\Delta R}{R}$ fittings.
			{\it Black:} using all our events, {\it blue:} using all our events with maximum pileup error below 20\%, {\it magenta:} using all our events with maximum pileup error above 20\%.
			{\bf Bottom:} Density structure (assuming fully-ionized plasma), as a function of $\Delta{h}$=$h-h_{ref}$ derived from the top plot (see text for details).
			Some data points at high densities have no error bars: this means the error bar was actually larger than the nominal value.
			We kept them on the plot because of their consistency.
			The squares represent the densities derived around 35 keV emission.
		}
		\label{fig:main}
		\end{figure}

	From the $\Delta{h}_i$ values obtained in the preceding paragraph, the plot in Figure~\ref{fig:main} (top) is constructed.
	It shows the altitude differences between emission at reference energy $\varepsilon_{ref}$=35 keV and emission at other energies (we have also looked at other choices of $\varepsilon_{ref}$, and found similar results, albeit with larger error bars when $\varepsilon_{ref}$ was higher than 35 keV).
	Also shown are curves derived from subsets of our data: {\it blue:} using only events with maximum pileup error $<$20\%, and {\it magenta:} using only events with maximum pileup error $>$20\%.
	Pileup issues will be discussed further on.
	The results presented in Figure~\ref{fig:main} (top) show a systematic decrease in altitude as a function of energy, despite somewhat largish error bars at high energies.
	Notice at low energies the departure from a strictly monotonic relationship.
	We attribute this to contamination of our sample by thermal emission at low energies. 

	We have checked the reliability of results in Figure~\ref{fig:main} (top) by varying certain selection criteria and redoing the procedure.
	We have tried to select events with quicklook images above 50 keV (as opposed to 25 keV), 
	tried other subcollimator combinations (e.g. 1--9), tried varying accumulation time intervals, different energy bands, and different $\varepsilon_{ref}$:
	the curve essentially remains the same, though generally with larger error bars.
	We have also verified that a histogram of the normalized residuals from the $\frac{\Delta{R}}{R}$ linear fitting are normally distributed, with mean 0 and standard deviation 1.

	The data presented in Figure~\ref{fig:main} (top) will be used in Section~\ref{sect:Derivation} to determine an average solar chromopheric/coronal density structure during flaring times.
	It can already be clearly seen that above $\sim$50 keV, flare footpoint centroids have a vertical extension of less than $\sim$0.5 Mm, 
	i.e. electrons above $\sim$50 keV reach their stopping heights within that 0.5 Mm region.
	We finally discuss the potentially important sources of contamination to our dataset in the next subsection.

	\subsection{Possible sources of contamination}\label{sect:contamination}
		\begin{itemize}
			\item {\it Presence of a thermal component:} At low energies (e.g. usually $\lesssim$15 keV for M-flares, and $\lesssim$25 keV for X-flares), 
				the non-thermal thick-target beam model can no longer be applied, due to the presence of high-temperature, high-altitude X-ray emitting loops.
				
			\item {\it Presence of a non-thermal coronal source:} \citep{Krucker2008b} These sources are often difficult to observe, due to instrumental dynamic range, but are probably present most of the time \citep{Krucker2008}.
				It is likely that such high-altitude source exist, slightly moving the center of gravity of our footpoint sources to higher projected altitudes.
				Assuming a coronal source altitude of 10 Mm above the footpoint(s), with 10\% of the footpoint non-thermal flux, the upward shift in altitude of the emission centroid could be $\sim$1 Mm.
				As coronal source spectra are much steeper than footpoint spectra \citep[e.g.][]{Marina2006}, this effect is less pronounced at higher energies.

			\item {\it Pulse pileup:} 
				At high photon fluxes, detector pulse-pileup occurs \citep{Smith2002}.
				This has the effect of combining two low-energy photon into a single higher-energy photon.
				Hence, an intense thermal loop at $\sim$18 keV will have a trace in $\sim$36 keV images (in the case of attenuator state 3).
				A pile-up error of 10\% (as given e.g. by the routine hsi\_pileup\_check.pro) under attenuator state 3, should roughly shift the center of gravity of 36 keV emission by $\sim$1 Mm to higher altitudes (assuming the thermal loop is about 10 Mm above the footpoints)
				This effect will typically occur only at high countrates, and is very energy-dependent, i.e. present mostly at twice the energy where most of the counts are, itself dependent on the spacecraft attenuator state:
				$\sim$12 keV for attenutator state 0, $\sim$24 keV for attenutator state 1, $\sim$36 keV for attenutator state 3.
				
			\item {\it Albedo:}
				Albedo effects \citep{Bai1978,Kontar2010} can also influence the results: up to 20--40\% of the 30--40 keV flux can be due to albedo (assuming isotropic beaming).
				Assuming most of the 30--40 keV emission comes from $\sim$2 Mm above the photosphere, a $\sim$30\% component reflected from the photosphere would shift the centroid of 30--40 keV emission by about $\sim$0.5 Mm downwards.
				This effect is present for all flares, is expected to be greatest around 30--40 keV, and shifts the centroid to lower altitudes, contrary to the pulse-pileup and coronal source effects.
				Furthermore, the amplitude of the effect of albedo is maximum near disc center, and minimum near the limb, and hence has a tendency to offset (in part) the effects of pulse pile-up and coronal sources.
		\end{itemize}

		To conclude, it can be said that observed heights for energies $\gtrsim$50 keV are trustworthy (within their statistical limitations),
		but that non-thermal emission $\lesssim$40 keV may be slightly offset (typically by up to 1--2 Mm) to higher altitudes than in reality,
		and with the magnitude of this offset varying somewhat with energy
		
		With these caveats in mind, we have attempted in the next section to derive an average density-height profile from our dataset.

\section{Derivation of densities}\label{sect:Derivation}

	\subsection{Direct derivation} \label{sect:direct}
		
		Figure~\ref{fig:main} (bottom) displays the density structure obtained using all the flares in our list, and using two subsets of it, with error bars.
		They were obtained using the method described in \citet{Brown2002}, and in Appendix~\ref{appendix:directderivation}.
		


		A fully-ionized corona was assumed, but high energy electrons (producing most of the high-energy radiation) probably reach regions of low ionization in the lower chromosphere.
		This means that densities obtained at low altitudes (derived from the higher energies) should actually be multiplied by $\sim$2.8 \citep[electron beams in a neutral plasma emits less HXR bremsstrahlung][]{Brown1973,Kontar2002}.
		
		Figure~\ref{fig:main} suggests that 
			1) emission at lowest energies are clearly contaminated by thermal loops, and results derived from those assuming non-thermal beam models are not to be trusted;
			2) emission above 50 keV is emitted from regions with densities $>$10$^{13}$ cm$^{-3}$, assuming fully-ionized plasma. 
			As we are probably in the chromospheric neutral region, the densities are likely $\sim$3 times as much.
			3) emission at intermediate energies ($\sim$20--40 keV) appear to come from regions of different densities, depending on the amount of pileup, itself dependent on the intensity of the non-thermal flux.
		From 2), one can hypothesize that $>$50 keV emission always comes from the same altitudes, whether we are dealing with small or large flares.
		This in turn tells us that emission at $\sim$35 keV for high pile-up events is almost 1 Mm above that of low pile-up events (top plot of Figure~\ref{fig:main}), 
		which is in accordance with the discussion in Section~\ref{sect:contamination}, but does not preclude the following scenario: 
		larger (and longer) flares, which typically have high pileup, are expected to have had greater chromospheric ablation by the time we reach the peak in non-thermal flux (when our observation are made),
		making the loop denser, and the emission at intermediate energies ($\sim$35 keV) higher. What we observe is an average of this effect over all our selected flares.
					
	Figure~\ref{fig:main} (bottom) suggests an atmosphere with density profile with at least two exponential components.
	In the next section, we have formulated simple density models and attempted to fit it to the data in Figure~\ref{fig:main}(top).

	\subsection{Density model fitting} \label{sect:fitting}
	
	We have attempted to fit our data with several density models, but will only discuss a two-exponential with unit step ionization (at altitude $h_{step}$: fully-ionized for altitudes $h>h_{step}$ and fully neutral for $h<h_{step}$) model, and only briefly mention some of the others.

	As in \citet{Kontar2008a}, we have added a data point for our fittings, which actually helps us in determining $h_{ref}$, our reference height for emission at $\varepsilon_{ref}$=35 keV:
	it is the well-established hydrogen density at the photosphere $n(h=0)=n_0$=1.16$\times$10$^{17}$ cm$^{-3}$.
	
	We have used the routine mpfit.pro (found e.g. in the IDL Astronomy Library) to make our fittings (and tried also IDL's amoeba.pro routine, with no discernible differences).
	We have used a Monte Carlo approach to determine error bars for our fitting parameters: for each fitting ``run'' (a hundred such runs were executed), a random amount was added to each data point.
	This amount is normally distributed, with mean 0, and standard deviation equal to the nominal error of the data value.

\begin{figure}[ht!]
		\centering
		\includegraphics[width=8cm]{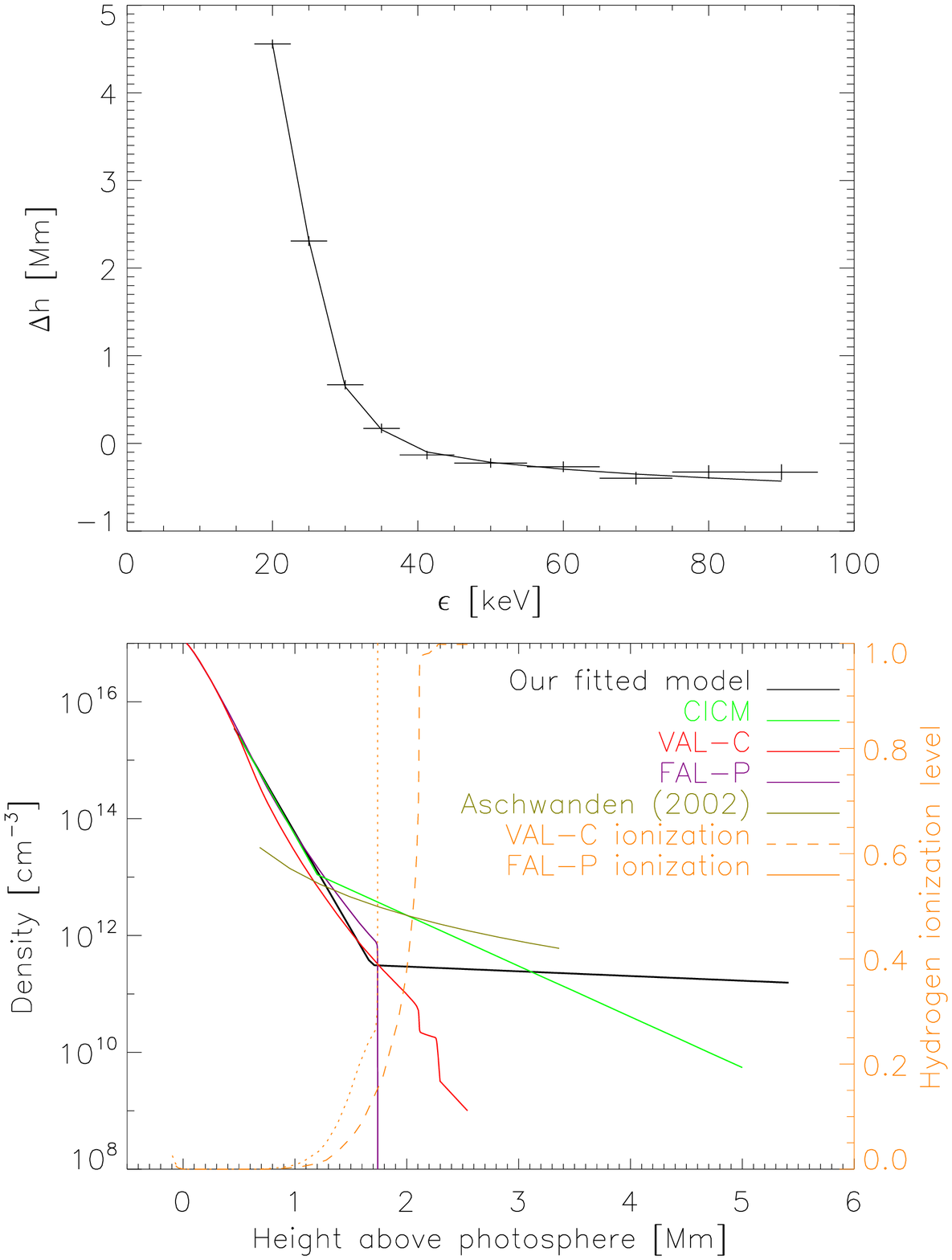}
		\caption{
			Two-barometric component atmosphere fitting on low-pileup data ($<$15\%), with unit step ionization variation.
			{\bf Top:}  Actual data points and fit (the horizontal ``error bars'' are actually binwidths).
			{\bf Bottom:} Fitted density structure {\it (solid black)}, 
				VAL-C atmospheric model \citep[][ {\it solid red}]{VAL1981}, 
				FAL-P atmospheric model \citep[][ {\it solid purple}]{Fontenla1993},
				CICM \citep[Caltech Irreference Chromosheric Model][ {\it solid green}]{Ewell1993},
				and \citet{Aschwanden2002} {\it (olive green)} results (which assumed a power-law density distribution and a fully-ionized atmosphere).
				{\it Solid orange:} FAL-P ionization level, {\it dashed orange:} VAL-C ionization level.
			}
		\label{fig:fit2exp_step}
		\end{figure}

	We show only the two-barometric component with a unit step ionization change model (Figure~\ref{fig:fit2exp_step}) for the low pileup case, 
	because we want to minimize the influence of pulse-pileup.
	Data with higher pileup and/or other more complicated models produced large $\chi^2$ results.
	We only fitted above $\sim$20 keV. Events with low pileup errors have typically little to no thermal emission above this threshold.
	
		\begin{table*}[ht!]
		\caption{Fitting parameters and other parameters derived from them. Double exponential structure, with unit step ionization, using flares with low pileup.}
		\centering
		\begin{tabular}{ccccc}
		\tableline\tableline
			Fitting parameters 	& Description		&	 Units	&	Value			\\
		\tableline
			$H_0$			& Low-altitude scale height	& km	&	131$\pm$16		\\
			$h_{01}$		& Altitude of transition between exponential components	& Mm	&	1.69$\pm$0.21		\\
			$H_1$			& High-altitude scale height	& Mm	&	5.4$\pm$0.6		\\
			$h_{step}$		& Altitude of unit step ionization change& Mm	&	1.3$\pm$0.2		\\
			$h_{acc}$		& Altitude of acceleration region	& Mm	&	19$\pm$32		\\
			$\chi^2$		&				& -	&	0.79			\\
		\tableline
		\tableline
			Derived			&				&	&				\\
		\tableline
			$n_{acc}$		& Density at acceleration altitude& cm$^{-3}$& $(8\pm71)\times10^8$	\\
			$\varepsilon_{step}$	& Initial electron energy required to reach neutral layer				& keV	&	36$\pm$6		\\
		\tableline
		\tableline

		\end{tabular}
		\label{tab:2exp_step}
		\end{table*}

	Table~\ref{tab:2exp_step} summarizes the result of the fittings. 
	The $h_{step}$ parameter is the altitude where the ionization abruptly changes from 100\% (corona) to 0\% (chromosphere), and $\varepsilon_{step}$ is the minimal initial energy required for electrons to reach the neutral layer.
	$h_{01}$ is the altitude of the change from the low-altitude exponential component to high-altitude exponential component.
	The scale heights $H_0$ and $H_1$ are trustworthy, thanks to the additional $n_0$ datapoint and to the tight error bars at lower energies.
	The acceleration altitude $h_{acc}$ and density $n_{acc}$ are unsurprisingly poorly constrained.
	The altitude of ionization change, $h_{step}$, is found to be $\approx$1.3$\pm$0.2 Mm above the photosphere, close to the 5\%--10\% ionization level in the VAL-C or FAL-P atmospheres (Figure~\ref{fig:fit2exp_step}).
	The low-altitude scale height was determined to be $\sim$131$\pm$16 km, which corresponds to a (neutral gas) temperature of 5600$\pm$700 K, consistent with chromopheric temperatures below the transition region.
	The high-altitude scale height of the flaring atmosphere is determined to be $\sim$5.4 Mm, which, assuming a fully-ionized isothermal plasma, corresponds to a temperature of $\sim$115 kK.
	As this region of the lower corona is very dynamic (i.e. non isothermal) during flares, and this value is an average over more than 800 flares, we suspect this usually mid-transition region temperature does not have any intrinsic value.

	%
	
	We would like to point out that events with higher maximum pulse pileup error required another exponential component at intermediate altitudes in order to produce a ``reasonable'' fit (albeit with reduced-$\chi^2 >$4).
	This led to a fit very close to the CICM and \citet{Aschwanden2002} results at intermediate altitudes (1.5--3 Mm), and, at low and high altitudes, similar to the shown low-pileup case.
	While an appealing result, we do not think it is trustworthy, given the afore-mentioned issues.
	
\section{Conclusion and future work} \label{sect:discussion}

	Using most of RHESSI's observations of flares with strong non-thermal HXR emission, we have derived an average emission energy vs. relative height profile.
	From this curve, it has been possible to derive densities, assuming a beam-like thick-target model.
	Furthermore, adding the well-known density at the surface of the photosphere as an additional data point,
	we could derive an average {\it absolute} height vs. density profile, and successfully fitted a two-exponential with step ionization model atmosphere to it.

	Although imaging show larger vertical spatial extent of a few Mm \citep{Kontar2010a}, we have found that flare footpoint {\it centroids} above $\sim$50 keV extent vertically, as expected from the thick-target beam model, and are within $\sim$0.5 Mm of each other, in regions with neutral densities well above 3$\times$10$^{13}$ cm$^{-3}$.
	We have found density scale heights at low altitudes/high energies of 131$\pm$16 km, matching well the VAL-C or FAL-P models \citep[as well as the 155$\pm$30 km value found by ][]{Kontar2010a}, and of 5.4$\pm$0.6 Mm at higher altitudes, for a flaring atmosphere.
	A host of other fitting parameters at intermediate heights or energies, and other derived quantities, could be obtained, but various intrumental, physical and/or statistical effects prevented us from obtaining reliable results for these.
	We have listed the different major sources of contamination and gave example of their possible contributions: thermal loops, non-thermal coronal source, pulse pileup, and albedo.
	The fitting results we presented minimized the impact of pulse-pileup, and were not affected by thermal contamination.
	
	While properly accounting for all these contaminating effects is difficult in a statistical study (particularly the contribution of any coronal source), we believe they are quantifiable and can be compensated for in a few well-chosen events.
	But in keeping with the statistical approach used so far, we plan on investigating the use of backprojection-based imaging with a very limited set of {\it fine} subcollimators:
	it is our hope that the lower sensitivity stemming from the use of a smaller number of collimators will be offset by the fact that taking the brightest pixel (as opposed to a centroid) in {\it spatially-resolved} sources should be less prone to the contaminants we have discussed so far.
	Our goal is to better resolve the intricate interplay between changes in density and changes in ionization level at intermediate altitudes ($\approx$1--3 Mm above photosphere, corresponding to intermediate energies of $\approx$ 25--50 keV).
	
\appendix
\section{Determining altitude differences from 2-D solar maps}\label{appendix:dh}

	A point in space has coordinates ($r$,$\theta$,$\phi$) in a spherical coordinate system centered on the Sun.
	Transforming into rectangular coordinates:
	\begin{eqnarray}
		x & = & r \sin\theta \cos\phi \\
		y & = & r \sin\theta \sin\phi \\
		z & = & r \cos\phi
	\end{eqnarray}

	Most solar maps use a planar coordinate system ($X$,$Y$) centered on the Sun, where $X$ and $Y$ represent angular distance from Sun center, as observed from Earth.
	Assuming that Earth is somewhere along the $x$ axis, we have:
	\begin{eqnarray}
		X \cdot d & = & y =  r \sin\theta\sin\phi	\\
		Y \cdot d & = & z =  r \cos\theta
	\end{eqnarray}
	Where $d$ is the Sun-Earth distance. 

	On a solar map, emission at energy $\varepsilon_i$ is located at position ($X_i$,$Y_i$).
	In this Appendix, we use lower case letters for real quantities, and upper case letters for quantities projected on the Sun.
	Assuming that emission at different energies is along a radial from the Sun, we have:
	\begin{eqnarray}
		X_i \cdot d & = & (r_s+h_i) \sin\theta\sin\phi	\\
		Y_i \cdot d & = & (r_s+h_i) \cos\theta	
	\end{eqnarray}
	where $r_s$ is the solar radius, and $h_i$ the altitude of emission at energy $\varepsilon_i$.
	
	Hence, between energies $\varepsilon_1$ and $\varepsilon_2$:
	\begin{eqnarray}
		\Delta X \cdot d & = & (X_2-X_1) \cdot d = (h_2-h_1) \sin\theta\sin\phi = \Delta{h} \sin\theta\sin\phi \\
		\overline{X} \cdot d & = & \frac{X_1+X_2}{2} \cdot d = \left( r_s + \frac{h_1+h_2}{2} \right) \sin\theta\sin\phi \approx r_s \sin\theta\sin\phi
	\end{eqnarray}
	And $\frac{\Delta{X}}{\overline{X}}$=$\frac{\Delta{h}}{r_s}$, or $\Delta{h}=r_s \frac{\Delta{X}}{\overline{X}}$. 
	The ratio $\frac{\Delta{X}}{\overline{X}}$ is obtained from non-thermal HXR flare maps and $r_s$ = 696 Mm, yielding $\Delta{h}$.
	This method is the one used by \cite{Matsushita1992}.

	Alternatively \citep[e.g.][]{Sato2006}:
	\begin{eqnarray}
		\Delta R \cdot d	& = & \sqrt{ (X_2-X_1)^2 + (Y_2-Y_1)^2 } = \Delta{h} \sqrt{ \sin^2\theta \sin^2\phi + \cos^2\theta } \\
		\overline{R} \cdot d	& = & \sqrt{ \overline{X}^2 + \overline{Y}^2 } \approx  r_s \sqrt{ \sin^2 \theta \sin^2 \phi + \cos^2 \theta }
	\end{eqnarray}
	And $\Delta{h}$=$r_s \frac{\Delta{R}}{\overline{R}}$, with $\frac{\Delta{R}}{\overline{R}}$ given by HXR flare maps.
	
	In this paper, we used this second method to determine $\Delta{h}$. 
	As it was using twice as much information, it was deemed the best. 
	Practically, both yielded very similar results.	

	See Figure~\ref{fig:rvsdr} for how the slopes $\frac{\Delta{R}}{\overline{R}}$ are derived from linear fitting of the data.

\section{Direct derivation of densities}\label{appendix:directderivation}

	As noted in \citet{Brown2002} \citep[see also][]{PSH2009a},
	and for electrons having a negative injected power-law distribution and propagating towards region of higher densities,
	non-thermal emission at energy $\varepsilon$ {\it is principally emitted by electrons near energy $\varepsilon$}, 
	and  {\it spatially peaks where electrons have crossed a column density: }
		\begin{equation}
			 N(s) \equiv \int_0^s n(s') ds' \approx \frac{\varepsilon^2}{2K}
		\end{equation}
	with $s$ the distance from the acceleration region, $\varepsilon$ in keV, and $K$=2.6$\times$10$^{-18}$ cm$^2$ keV$^2$ (fully-ionized corona).
	It is important to note \citep[as amply discussed in][]{Brown2002} that $N(s) \approx \frac{\varepsilon^2}{2K}$ is an approximation,
	and that in truth a dependence on the spectrum of the injected electrons and on the density structure exists \citep[see also][]{Xu2008,Prato2009}.

	Using the previous approximation, and assuming that $s$ is radial and using instead the altitude $z$ as variable, the density $n(z)$ is simply obtained through derivation:
	\begin{equation}
		n(z)=-\frac{dN(z)}{dz}=-\frac{\varepsilon}{K}\frac{d\varepsilon}{dz}
	\end{equation}
	(In a more general treatment, one can account for the changing ionization level with altitude by making $K$ a function of $z$.)

	For discrete data points, where $z_i$ are the position of maximal emission at energy $\varepsilon_i$, and for i$\neq$j:
	\begin{equation}
		n(z_{ij}=\frac{z_i+z_j}{2}) \approx -\frac{(\varepsilon_i+\varepsilon_j)}{2K}\frac{\varepsilon_j-\varepsilon_i}{z_j-z_i} = \frac{1}{2K} \frac{ \varepsilon_j^2 - \varepsilon_i^2}{z_i-z_j}
	\end{equation}
	Our data yields $\varepsilon_i$ and $\Delta{h}_i$=$z_i-z_{ref}$.
	$z_{ij}$=$h_{ref}$+$\frac{\Delta{h}_i + \Delta{h}_j}{2}$, with $h_{ref}$ to be determined by other means.
	
\section{Model fitting}\label{appendix:modelfitting}
	
	\subsection{Single exponential with uniform ionization} \label{appendix:modelfitting:1}
		This simple analytical model will not be used, but is shown as a starting point for other models.
		
		Ionization level is assumed 100\% throughout.
		The density structure is modeled using:
		\begin{equation} \label{EQ:MF1}
			n(h) = n_0 e^{-h/H} = n_{acc} e^{-(h-h_{acc})/H}
		\end{equation}
		with $n_0=1.16 \times 10^{17}$ cm$^{-3}$, the well-accepted photospheric value, $h$ is the altitude above photosphere, 
		$H$ is the scale height, and $h_{acc}$ and $n_{acc}$ the height and density of the acceleration region. 
		
		The column density is:
		\begin{equation}
			N(h) = - \int_{h_{acc}}^h n(h) dh = n_0 H \left( e^{-h/H} - e^{-h_{acc}/H} \right) = n_{acc} H \left( e^{-(h-h_{acc})/H} - 1 \right)
		\end{equation}
		where $n_{acc}$ is the density in the acceleration region.
		
		Using the $\varepsilon^2 \approx 2KN$ approximation, we get:
		\begin{equation} \label{EQ:MF2}
			\varepsilon^2 = 2 K n_{acc} \left(  e^{-(h-h_{acc})/H} -1 \right)
		\end{equation}
		and:		
		\begin{equation} \label{EQ:MF3}
			h = h(\varepsilon) = -H \ln \left( 1 + \frac{\varepsilon^2}{2KHn_{acc}} \right)
		\end{equation}
		
		And the difference in height $\Delta h_{ij}$ between the centroid of emission at energies $\varepsilon_i$ and $\varepsilon_j$:
		\begin{equation}
			\Delta h_{ij} = h_i - h_j = H \ln \left(  \frac{1+\frac{\varepsilon_{j}^2}{2KHn_{acc}}}{1+\frac{\varepsilon_{i}^2}{2KHn_{acc}}} \right)
		\end{equation}
	
		The data yield $\Delta h_{ij}$, $\varepsilon_i$ and $\varepsilon_j$, which can be used to determine the fitting parameters $H$ and $n_{acc}$.
		And, finally, $h_{ref}$ can be determined using Eq. \ref{EQ:MF3} with $\varepsilon$=$\varepsilon_{ref}$.
		Notice that only good knowledge of $H$ is required to determine $h_{ref}$ with accuracy, and is independent of what $n_{acc}$ (or, alternatively, $h_{acc}$) is.
		As we will see, this will prove to be very useful, as $n_{acc}$ or $h_{acc}$ will often turn out to have quite large errors.
	
	\subsection{Double exponential with uniform ionization} \label{appendix:modelfitting:2}
		Ionization level is assumed 100\% throughout. The density structure $n(h)$ has $h$ as variable and $H_0$, $H_1$, and $h_{01}$ as parameters.
		\begin{eqnarray} \label{eq:1}
			n(h)	& = &	\left\{	\begin{array}{ll} 
						n_0 e^{-h/H_0}			& ,  \,\,\, h < h_{01}	\\
						n_{01} e^{-(h-h_{01})/H_1}	& ,  \,\,\, h > h_{01}  \\
						\end{array} \right.
		\end{eqnarray}
		with:
		\begin{eqnarray}
			n_0	& = &	1.16 \times 10^{17}	cm^{-3}	\\
			n_{01}	& = &	n_0 e^{-h_{01}/H_0}			\\
			n_{acc}	& = &	n_{01}	e^{-(h_{acc}-h_{01})/H_1}	\\
		\end{eqnarray}

		The column density $N(h)$ requires an additional parameter, $h_{acc}$, the height of the acceleration region.
		\begin{equation}
			N(h)	=	- \int_{h_{acc}}^h n(h) dh
		\end{equation}
		which can easily be integrated to:
		\begin{eqnarray}
			N(h)	& = &	\left\{	\begin{array}{ll} 
						0											& ,  \,\,\, h > h_{acc}\\
						H_1 n_{acc} (e^{-(h-h_{acc})/H_1}-1)							& ,  \,\,\, h_{01} < h < h_{acc}\\
						H_1 n_{acc} (e^{-(h_{01}-h_{acc})/H_1}-1) + H_0 n_{01} (e^{-(h-h_{01})/H_0}-1)		& ,  \,\,\, h < h_{01}	\\
						\end{array} \right.
		\end{eqnarray}
		
		The rest of the procedure is similar to as explained in Section \ref{appendix:modelfitting:1}, except for the following:
		(a) We have two additional fitting parameters $h_{01}$ and $H_1$. 
		(b) The explicit $\varepsilon(h)$ expression that we get is no longer easily invertible into a simple $h(\varepsilon)$ equation, as done in \ref{EQ:MF2} and \ref{EQ:MF3}.
		The inversion is hence done numerically, using interpolations.						

	\subsection{Double exponential with unit step ionization} \label{appendix:modelfitting:3}
		
		We use the same equations as in the previous section, except we add another parameter, $h_{step}$, the altitude at which the solar atmopshere is assumed to abruptly change from fully ionized to fully neutral \citep[][and citations therein]{Kontar2002}.
		In the computation of $N(h)$, the density in regions below $h_{step}$ is weighted by a factor 1/2.818 \citep{Kontar2002}.

	\subsection{Triple exponential with uniform ionization, or with unit step ionization change} \label{appendix:modelfitting:4}
		As in \ref{appendix:modelfitting:2} and \ref{appendix:modelfitting:3}, with an additional intermediate exponential component.

\section{Analytical proof that $\frac{d^2I}{dN^2}<0$, $\forall N$ and $\forall \delta>2$, with consequence that for an injected negative power-law ($\delta>2$) distribution of electrons propagating under the thick-target assumption
to have peaks of increasing energy emission as it propagates, the density gradient along the beam's path must be positive:}\label{appendix:proof}
	
	From \citet{Brown2002,PSH2009a}, the emissivity per unit column density of such a distribution of electron is:
		\begin{equation}
			\frac{dI}{dN} = \frac{A}{\varepsilon} (2KN)^{-\delta/2} \beta \left( \frac{1}{1+u};\frac{\delta}{2};\frac{1}{2} \right),
		\end{equation}
	where $\varepsilon$ is the emitted photon energy, $N$ the column density crossed by the propagating electrons, $\delta$ the negative spectral index of 
	the injected electrons, $\beta$ the incomplete Beta function, $u=\frac{\varepsilon^2}{2KN}$, and the constants $K$=2.6$\times$10$^{-18}$ cm$^{2}$ keV$^{-2}$, 
	and $A$=$(\delta-1)\frac{F_1}{E_1} \overline{z^2} \frac{\kappa_{BH}}{8\pi D^2} E_1^{\delta}$ \citep[see e.g.][for a complete explanation of these constants]{Brown2002}.
	$\frac{dI}{dN}$ is $\propto \varepsilon^{-\delta-1}$ (flat with $N$) when $\frac{\varepsilon^2}{2KN} \gg 1$, and is $\propto \varepsilon^{-1} N^{-\delta/2}$ when $\frac{\varepsilon^2}{2KN} \ll 1$ \citep[see Figure~\ref{fig:d2IdN2} {\it top} and e.g.][, Appendix A]{PSH2009a}.
	Straightforward derivation yields:
		\begin{equation}
			\frac{d^2I}{dN^2} = \frac{A}{\varepsilon} 2K (2KN)^{-\delta/2-1} \left[ u^{1/2}(1+u)^{-\delta/2-1/2} - \frac{\delta}{2} \beta \left( \frac{1}{1+u};\frac{\delta}{2};\frac{1}{2} \right)  \right]
		\end{equation}
	which happens to have the shape shown in Figure~\ref{fig:d2IdN2} {\it (bottom)}:
		\begin{figure}[ht!]
		\centering
		\includegraphics[width=7cm]{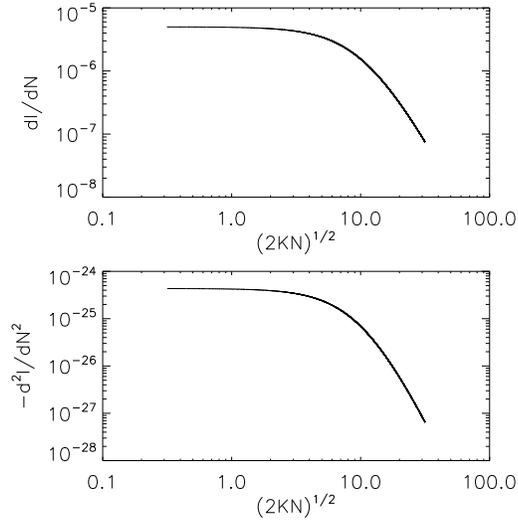}
		\caption{
			Shape of $\frac{dI}{dN}$ and $-\frac{d^2I}{dN^2}$ (both in arbitrary units) as a function of $\sqrt{2KN}$, the latest corresponding to the minimal initial electron energy required to cross column density $N$.
		}
		\label{fig:d2IdN2}
		\end{figure}
	Notice that $\frac{d^2I}{dN^2}<$0 for all $\delta$, $\varepsilon$, and $N$ of interest. 
	The mathematical proof, valid for $\delta>2$, is as follows: 
	$\frac{d^2I}{dN^2}<0$ if $\frac{\delta}{2} \beta \left( \frac{1}{1+u};\frac{\delta}{2};\frac{1}{2} \right) > u^{1/2}(1+u)^{-\delta/2-1/2}$, 
	i.e. if $f(u,\delta)>1$, with $f(u,\delta) = \frac{\delta}{2} \beta \left( \frac{1}{1+u};\frac{\delta}{2};\frac{1}{2} \right) u^{-1/2}(1+u)^{\delta/2+1/2}$.
	Note that $\beta \left( \frac{1}{1+u};\frac{\delta}{2};\frac{1}{2} \right)$ = $\int_0^{\frac{1}{1+u}}x^{\delta/2-1}(1-x)^{-1/2}dx$, which is $> \int_0^{\frac{1}{1+u}}x^{\delta/2-1}dx$, as $x \in [0,1[$.
	The latter can be integrated to $\frac{2}{\delta} (1+u)^{-\delta/2}$, assuming $\delta>2$ (a requirement with most thick-target formulae, and supported observationally).
	Hence $\beta \left( \frac{1}{1+u};\frac{\delta}{2};\frac{1}{2} \right) > \frac{2}{\delta} (1+u)^{-\delta/2}$, leading to $f(u,\delta) > \sqrt{1+\frac{1}{u}}$.
	Hence $f(u,\delta) > 1$ for all $u$ of interest ($u \in \, ]0,\infty[$). QED.

	For a localized peak in emission $\varepsilon$ to exist, we must have:
	\begin{eqnarray}
		\frac{d^2I}{dz^2} & \equiv & n \frac{d}{dN} \left( n \frac{dI}{dN} \right) =  n \left( \frac{dn}{dN} \frac{dI}{dN} + n \frac{d^2I}{dN^2} \right)	\\
				  & = & 0 
	\end{eqnarray}
	As $\frac{dI}{dN}>0$, $n>0$, and $\frac{d^2I}{dN^2}<0$, this condition can only be achieved if $\frac{dn}{dN}>0$, 
	i.e. when the density increases as the column density traversed increases, i,e. when our beam of electrons propagates towards regions of higher densities.

\bibliographystyle{apj}
\bibliography{psh_biblio}



\acknowledgments

We would like to thank Gordon Hurford for numerous discussions, and of course for making RHESSI visibilities a reality.
We would also like to thank the referee for very pointed comments.
This work was supported by NASA Contract No. NAS 5-98033 and NASA Heliospheric Guest Investigator grant NN07AH74G.
R. Lin has been supported in part by the WCU grant (No. R31-10016) funded by the Korean Ministry of Education, Science and Technology.



{\it Facilities:} \facility{RHESSI}.



\end{document}